\documentclass[aps,prd,preprint,floats,epsf,superscriptaddress,nofootinbib,amsmath]{revtex4}
\usepackage{graphicx}
\usepackage{amssymb,amsmath}
\usepackage{mathrsfs}
\usepackage{slashed}
\usepackage{indentfirst}
\usepackage{graphicx}
\usepackage{xcolor}
\usepackage{dsfont}
\usepackage{ulem}

\begin{document}


\bigskip

\title{Explaining the DAMPE $e^+ e^-$ excess using the Higgs triplet model with a vector dark matter}

\author{Chuan-Hung Chen}
\email[e-mail: ]{physchen@mail.ncku.edu.tw}
\affiliation{Department of Physics, National Cheng-Kung University, Tainan, Taiwan 70101, R.O.C.}

\author{Cheng-Wei Chiang}
\email[e-mail: ]{chengwei@phys.ntu.edu.tw}
\affiliation{Department of Physics, National Taiwan University, Taipei, Taiwan 10617, R.O.C.}
\affiliation{Institute of Physics, Academia Sinica, Taipei, Taiwan 11529, R.O.C.}

\author{Takaaki Nomura}
\email[e-mail: ]{nomura@kias.re.kr}
\affiliation{School of Physics, KIAS, Seoul 130-722, Korea}

\date{\today}

\begin{abstract}
We explain the $e^+ e^-$ excess observed by the DAMPE Collaboration using a dark matter model based upon the Higgs triplet model and an additional hidden $SU(2)_X$ gauge symmetry.  Two of the $SU(2)_X$ gauge bosons are stable due to a residual discrete symmetry and serve as the dark matter candidate.  We search the parameter space for regions that can explain the observed relic abundance, and compute the flux of $e^+ e^-$ coming from a nearby dark matter subhalo.  With the inclusion of background cosmic rays, we show that the model can render a good fit to the entire energy spectrum covering the AMS-02, Fermi-LAT, CALET and DAMPE data.
\end{abstract}


\maketitle


{\it Introduction --}
In a recent report, the Dark Matter Particle Explorer (DAMPE) Collaboration showed the high-energy cosmic-ray electron-and-positron (CRE) spectrum from 25~GeV to 4.6~TeV with unprecedented precision and power to discriminate between electrons and protons~\cite{Ambrosi:2017wek}.  The overall data can be well fit with a smoothly broken power-law model in the range of 55~GeV to 2.63~TeV.  However, a peak in the bin of $\sim 1.5$~TeV stands out of the continuum.  Though statistically insignificant yet, such a sharp excess of electrons and positrons in the cosmic rays still leads people to wonder that they may come from the annihilation of dark matter (DM) in a nearby subhalo.  In particular, quote a few works propose DM models involving a new leptophilic interaction~\cite{Yuan:2017ysv,Fan:2017sor,Gu:2017gle,Duan:2017pkq,Zu:2017dzm,Tang:2017lfb,Chao:2017yjg,Gu:2017bdw,Cao:2017ydw,Duan:2017qwj,Liu:2017rgs}.  A model-independent analysis regarding which type of DM model to fit data and constraints better is given in Ref.~\cite{Athron:2017drj}.

In this work, we revisit a DM model~\cite{Chen:2015cqa} proposed a few years ago.  (See also similar models in Refs.~\cite{Dev:2013hka,Chen:2014lla}.)  In the model, we extend the Higgs triplet model (HTM)~\cite{Konetschny:1977bn,Schechter:1980gr,Cheng:1980qt} with a hidden gauge symmetry of $SU(2)_X$ that is broken to its $Z_3$ subgroup by a quadruplet scalar field.  Such a symmetry breaking scheme renders the new gauge bosons stable and good candidates for weakly-interacting massive particle (WIMP) DM.  By coupling the complex Higgs triplet field and the $SU(2)_X$ quadruplet scalar field, the vector DM bosons can annihilate through one component of the $SU(2)_X$ quadruplet into a pair of doubly-charged Higgs bosons, each of which in turn decays into like-sign leptons provided the Higgs triplet vacuum expectation value (VEV) is sufficiently small.  Assuming the like-sign electrons and positrons as the dominant decay modes of the doubly-charged Higgs boson, we show that the required excess at $\sim 1.5$~TeV in the DAMPE CRE spectrum can be produced.  Besides, the model provides a link between neutrino mass and dark matter phenomenology.

{\it The model --}
Since the model has been detailed in Ref.~\cite{Chen:2015cqa}, here we only review the relevant parts for explaining the DAMPE CRE excess.  In addition to the SM gauge group, the model has an additional $SU(2)_X$ symmetry with the associated gauge field and coupling strength denoted by $X_\mu^a$ and $g_X$, respectively.  The $SU(2)_X$ symmetry is broken by a quadruplet field $\Phi_4= (\phi_{3/2}, \phi_{1/2}, -\phi_{-1/2}, \phi_{-3/2})^T/\sqrt{2}$ that does not carry SM gauge charges, where the subscript stands for the eigenvalue of the third generator (denoted by $\tau_3$) for the field and we use the phase convention that $\phi_{-i} = \phi^*_{i}$.  As in the HTM, we have a complex Higgs field $\Delta$ that is a triplet under the SM $SU(2)_L$ and carries hypercharge $Y = 1$, where we adopt the convention that the electric charge $Q = T_3 + Y$.

By requiring that the $\Phi_4$ field spontaneously develops a VEV in the $\tau_3 = \pm 3/2$ component:~\cite{Chen:2015nea}
\begin{align}
\phi_{\pm 3/2} = \frac{1}{\sqrt{2}} \left( v_4 + \phi_r \pm i \xi \right) 
\label{eq:vev}
\end{align}
with $v_4 \sim {\cal O}(10)$~TeV, the $SU(2)_X$ symmetry is broken and the gauge bosons $\chi_\mu ~(\bar \chi_\mu)= (X^1_\mu \mp i X^2_\mu)/\sqrt{2}$ and $X^3_\mu$ acquire mass: $m_{\chi} =\sqrt{3} g_X v_4 /2$ and $m_{X^3}=\sqrt{3} m_\chi$, all at the TeV scale.  The model still has a residual $Z_3$ symmetry, under which $\chi_\mu$ and $\bar \chi_\mu$ carry nonzero charges.  Such a discrete gauge symmetry ensures the stability of $\chi_\mu$ and $\bar \chi_\mu$ to be good DM candidates.  Here $\phi_r$ plays the role of a messenger between the hidden sector and the visible sector through the gauge interaction given by~\cite{Chen:2015nea}
\begin{align}
I_{\chi \bar\chi \phi_r} =& \sqrt{3} g_X m_\chi \phi_r \chi_\mu \bar\chi^\mu ~.
\label{eq:igg}
\end{align}

After the electroweak symmetry breaking as in the SM, the Higgs triplet is induced to develop a VEV, serving as a source of Majorana mass for neutrinos.  We parameterize the Higgs doublet and triplet fields as
\begin{align}
\Phi = 
\begin{pmatrix}
G^+ \\ \frac{1}{\sqrt2}(v+\phi + i G^0)
\end{pmatrix}
~~\mbox{and}~~
\Delta = \begin{pmatrix}
    \delta^+/\sqrt{2} & \delta^{++}  \\ 
    (v_\Delta + \delta^0 + i\eta^0)/\sqrt{2} & -\delta^+/\sqrt{2} \\   
\end{pmatrix} ~,
\end{align}
where the triplet VEV $v_\Delta$ is constrained by the electroweak rho parameter to be less than a few GeV~\cite{Cheng:1980qt,Gunion:1990dt,Patrignani:2016xqp}.  To produce the CRE excess given by the DAMPE experiment, we assume that the dominant decay modes of the charged Higgs boson to be leptonic.  In this case, $v_\Delta$ is required to be less than $\sim 10^{-4}$~GeV\,
\footnote{In general, the doubly-charged (singly-charged) Higgs boson can also decay to $W^\pm W^\pm (W^\pm h, W^\pm Z, t b)$ mode(s) whose widths are proportional to $v_\Delta^2$.  The branching ratios of these modes are suppressed when $v_\Delta \lesssim 10^{-4}$~GeV, and the only dominant decay modes are leptonic~\cite{Perez:2008ha}.}.

Because of the hierarchy of the VEV's among $\langle \Phi_4 \rangle$, $\langle \Phi \rangle$ and $\langle \Delta \rangle$: $v_4 \gg v \gg v_\Delta$, there is little mixing among $\phi_r$, $\delta^0$ and $\phi$.  We will therefore identify them as the physical Higgs bosons $H$, $\delta^0$ and $h$, respectively, with the masses:
\begin{align}
&
m_h \approx  m_\phi = \sqrt{2 \lambda}\, v ~,~~~ 
m_{H} \approx m_{\phi_r}= \sqrt{2 \lambda_\Phi}\, v_4 ~,
\nonumber \\
&
m_{\delta^{\pm\pm} } \approx m_{\delta^{\pm}} \approx m_{\delta^0}= m_{\eta^0} 
\equiv m_\Delta ~.
\end{align}
In the limit of no mixing, $H$ does not couple with the SM particles directly.  Instead, it interacts with the visible sector via the interactions described by~\cite{Chen:2015nea}
\begin{align}
I_{H \Delta \bar\Delta} = 
v_4 \lambda_4 H \left[ \delta^{++} \delta^{--} + \delta^{+} \delta^{-} 
+ \frac{1}{2} ( \delta^{0^2} + \eta^{0^2}) \right] ~.
\label{eq:Hdd}
\end{align}
To explain the DAMPE $e^+e^-$ excess, we take $m_\chi \approx 3$~TeV, $m_{\Delta} / m_\chi = 0.995$, and $m_{H} \approx 6$~TeV.  Such a parameter choice is consistent with the latest collider search of the doubly-charged Higgs boson that gives a lower bound of $m_\Delta > 770 - 800$~GeV~\cite{ATLAS:2017iqw}.

{\it Dark matter relic abundance and annihilation signal --}
Here we consider the scenario where the DM particles annihilate through the $\chi$-$\bar\chi$-$H$ gauge interaction given in Eq.~\eqref{eq:igg} and the $H$-$\Delta$-$\bar\Delta$ interaction given in Eq.~\eqref{eq:Hdd} into a pair of Higgs triplet bosons.  The assumed masses above result in the Breit-Wigner resonance enhancement in the pair annihilation process.  This in turn affects both relic abundance of DM~\cite{Gondolo:1990dk} and positron/antiproton fluxes~\cite{Feldman:2008xs,Ibe:2008ye}.  Subsequently, the Higgs triplet bosons decay dominantly into leptonic final states: $\delta^{\pm\pm} \to \ell^\pm{\ell'}^\pm$ and $\delta^\pm \to \ell^\pm \nu_{\ell'}$ with details depending on the values of $v_\Delta$ and the lepton Yukawa couplings with $\Delta$.  For definiteness, we assume that ${\cal B}(\delta^{\pm\pm} \to e^\pm{e}^\pm) \approx {\cal B}(\delta^\pm \to e^\pm \nu_{e}) \approx 100\%$\,
\footnote{ This assumption is not crucial in our analysis.  All that is required is that the two modes are dominant in the decays of singly- and doubly-charged Higgs bosons.  This occurs when neutrinos have an inverted mass hierarchy and nonzero Majorana phases in mixing~\cite{Perez:2008ha}.}.

In the non-relativistic limit, the DM annihilation cross section is given by~\cite{Chen:2015nea}
\begin{align}
 \sigma v \simeq
\frac{1}{192\pi} \left( \frac{\lambda_4}{m_\chi} \right)^2
\left[  
\left( \frac{v^2}{4} + 2 \epsilon \right)^2
+ \frac{\Gamma_H^2}{4m_\chi^2}
\right]^{-1}
\sqrt{1 - \frac{m_\Delta^2}{m_\chi^2}} ~,
\label{eq:sigmav}
\end{align}
where $\epsilon \equiv 1 - \frac{m_H}{2m\chi}$ and $\Gamma_H$ denotes the total width of $H$.  $\Gamma_H$ is found to be much smaller than $m_\chi$ and, therefore, its effect can be neglected.  The average speed of DM $v$ in units of the speed of light is typically $\sim 10^{-3}$ at the current Universe and $\sim 0.3$ at the freeze-out.  After fixing $m_\chi = 3$~TeV and $m_\Delta = 0.995 m_\chi$, $\sigma v$ is seen to depend only on two parameters: $\lambda_4$ and $m_H$.

\begin{figure}[t!]
\centering 
\includegraphics[width=8.4cm]{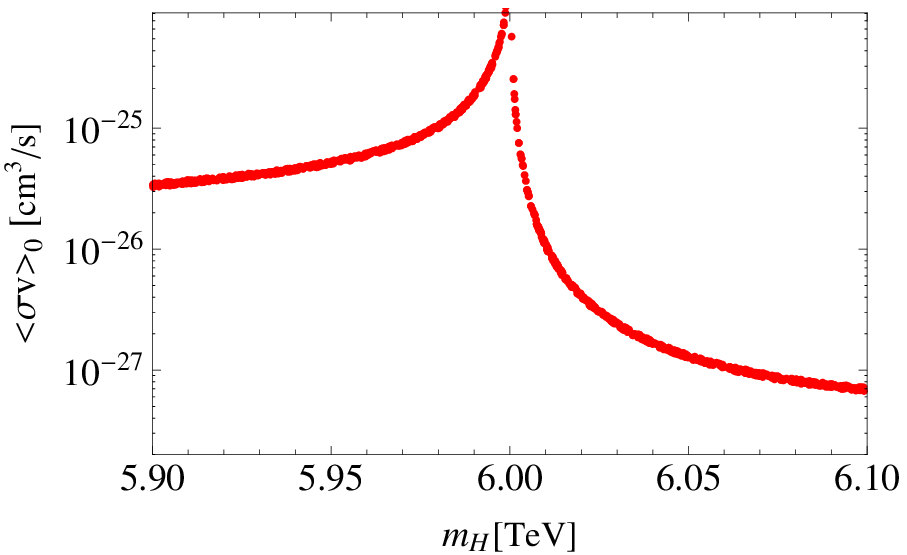}%
\includegraphics[width=8cm]{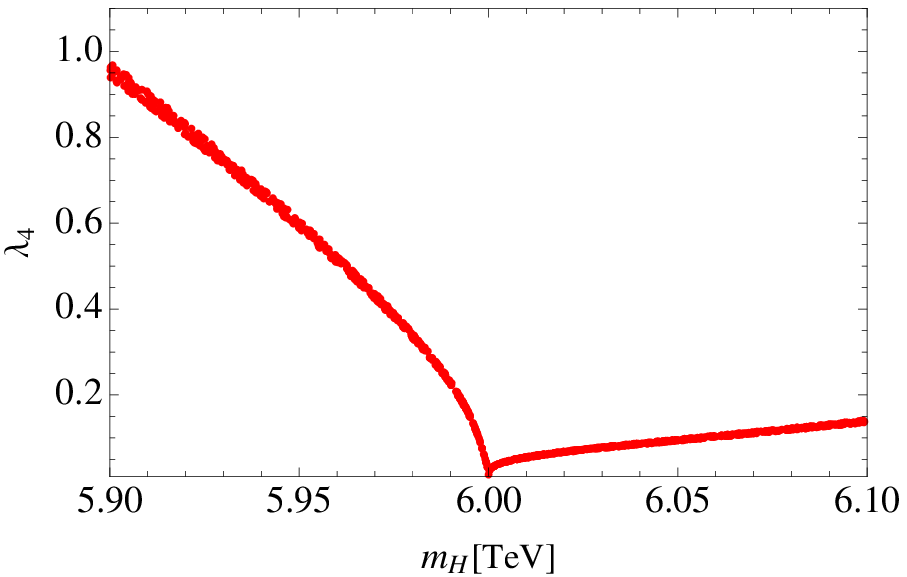}%
\caption{Parameters required to fit the DM relic abundance in the vicinity of the Breit-Wigner enhancement region.  Left plot: thermally averaged DM annihilation cross section as a function of $m_H$.  Right plot: The quartic coupling $\lambda_4$ as a function of $m_H$.  In these plots, we have fixed the DM mass $m_\chi = 3$~TeV.}
\label{fig:range}
\end{figure}

Our numerical analysis is done by utilizing the {\tt micrOMEGAs 4.3.5} package~\cite{Belanger:2014vza} implemented with the model to solve the Boltzmann equation for the observed cold DM relic abundance, $\Omega_C h^2 = 0.1199 \pm 0.0027$~\cite{Ade:2013zuv}, at the $2\sigma$ confidential level (CL).  We show in the left plot of Fig.~\ref{fig:range} the thermally averaged DM cross section at the current Universe.  It has a typical value around $10^{-27} - 10^{-25}$~cm$^3/$s around $m_H = 6$~TeV.  The right plot gives the relation between $\lambda_4$ and $m_H$.  The coupling $\lambda_4$ is seen to be perturbative within the displayed mass range of $m_H$.  The peculiar behavior at $m_H = 6$~TeV in both plots is owing to the resonance effect in the s-channel DM annihilation process.

{\it CRE spectrum --}
As alluded to before, we assume that the CRE excess observed by the DAMPE experiment comes from nearby subhalo,
where  the source term $q^c_{\rm DM}(\vec{x},E)$ in the diffusion-loss equation for a nearby DM clump at $x=x_c$ is given by~\cite{Brun:2009aj}:
\begin{align}
q^c_{\rm DM}(\vec{x},E) &= \frac{1}{4}  \langle \sigma v \rangle_0 \frac{L}{ m_{\chi}^2} \delta^3(\vec{x} -\vec{x}_c)
~~\mbox{with}~
L=\int \rho^2 dV \,,
\end{align}
where $\langle \sigma v \rangle_0$ is the current thermally averaged cross section, $L$ denotes the subhalo luminosity, and $\rho$ is the DM density profile. In the following, we consider the expected CRE flux by taking into account propagation effects.  The differential flux of $e^\pm$, defined as $d\Phi_{e^\pm}/dE = c F/4\pi$ in any point of our Galaxy, is given by~\cite{Cirelli:2010xx}:
\begin{align}
\frac{d\Phi_{e^\pm}}{dE} &= \frac{c}{4\pi b(E,\vec{x})} \left[    L \int^{m_{\chi}}_{E} dE_s \sum_f \left(  \kappa_f  \frac{dN^f_{\pm}}{dE}(E_s) \right)  \frac{I(E,E_s,\vec{x} )}{(4\pi \lambda(E,E_s,\vec{x}))^{3/2}}  \right]
~,
\label{eq:diffflux}
\end{align}
where the speed of $e^\pm$ is approximated by the speed of light, $v_{e^\pm} = c$, the effect through an annihilation channel $f$ is expressed as $\kappa_f =\langle \sigma v\rangle^f_0/(4 m_\chi)$, $E_s$ denotes the $e^\pm$ energy at the production,  $E$ is the observed $e^\pm$ energy, $b(E,\vec{x})$ is the $e^\pm$ energy loss coefficient function, and $I(E,E_s,\vec{x})$ is the generalized halo function.  The detailed definition of parameter $\lambda(E,E_s,\vec{x})$ can be found in Ref.~\cite{Kuhlen:2009is}, where it is not only related to the diffusion coefficient function and $b(E,\vec{x})$, but also to the propagation distance of $e^\pm$. Representing the Green function of the diffusion-loss equation with the electrons and positrons produced by the DM annihilation as the source, the generalized halo function is independent of the DM model.

\begin{figure}[tbh]
\centering 
\includegraphics[width=9cm]{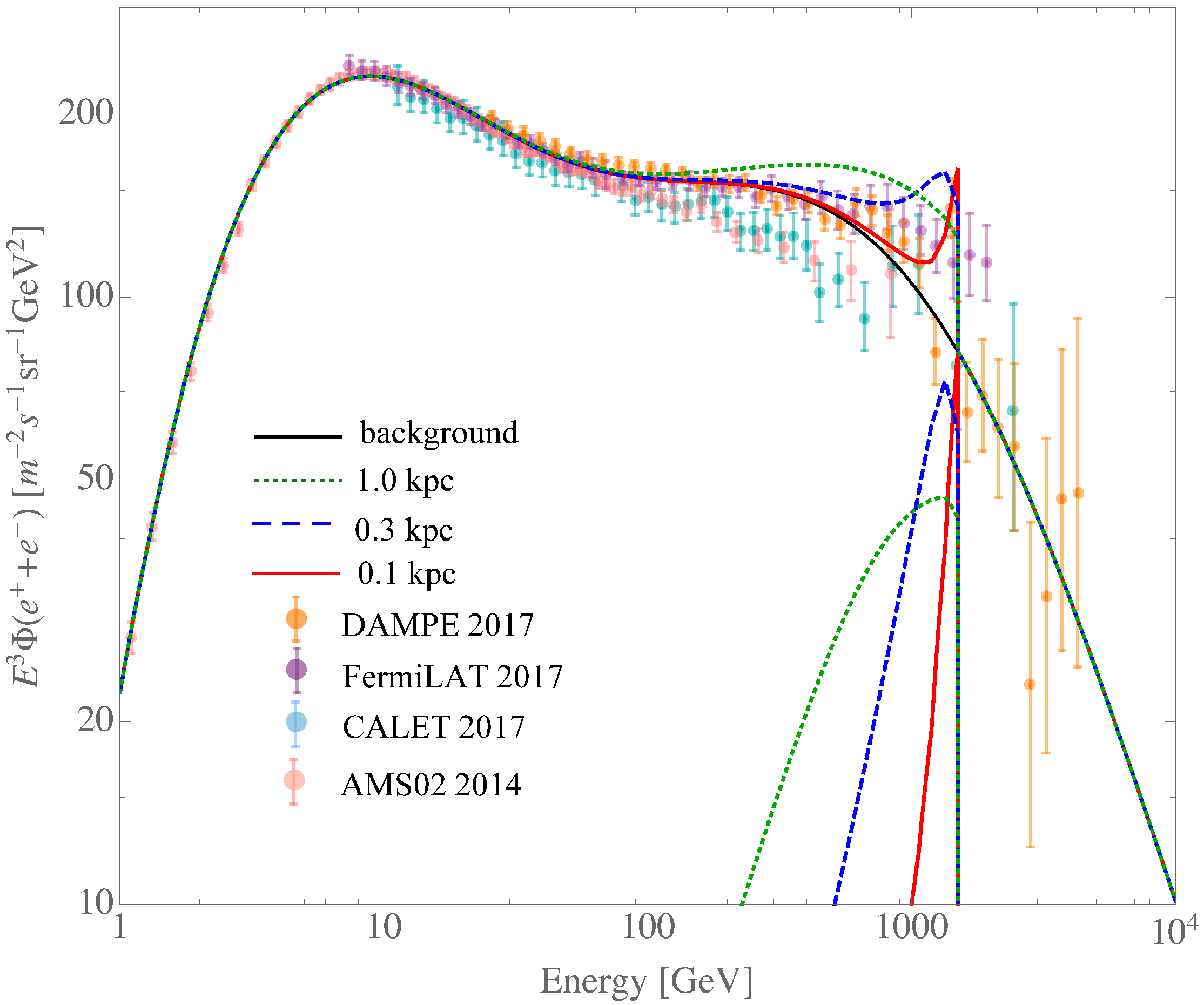}
\caption{The CRE spectrum (scaled by $E^3$) in the range of $1-10^4$~GeV.  The AMS-02 data are drawn in pink, the Fermi-LAT data in purple, the CALET data in cyan, and the DAMPE data in orange.  The black curve is the background contribution.  The excess from DM annihilation in our model is shown for three scenarios: the red curve for a subhalo at a distance of $0.1$~kpc; the blue dashed curve for $0.3$~kpc; and the green dotted curve for $1$~kpc.}
\label{fig:spectrum}
\end{figure}

In addition to the peak at around $E\sim 1.5$~TeV, to obtain $e^+ e^-$ spectrum from GeV to TeV, we also need the fluxes of cosmic-ray electron and positron backgrounds, which arise from various astrophysical sources.  We follow the parametrizations in Ref.~\cite{Ibarra:2013zia}, in which the results were used to fit the AMS-02 data~\cite{Aguilar:2013qda}.  For positrons, we write the flux as:
\begin{align}
\Phi^{\rm sec,IS}_{e^+}(E) & = C_{e^+} E^{-\gamma_{e^+}}\,, \  
\Phi^{\rm source,IS}_{e^+}(E)  = C_{s} E^{-\gamma_{s}} \exp(-E/E_s)\,,
\end{align}
 where $\Phi^{\rm sec,IS}_{^+}$ and $\Phi^{\rm source,IS}_{^+}$ denote the secondary and primary interstellar (IS) positrons, respectively, and the free parameters $C_{e^+, s}$, $\gamma_{e^+ ,s}$, and $E_s$ are determined by the cosmic-ray data.   Thus, the spectrum at the top of atmosphere (TOA) is given by~\cite{Perko:1987gky}:
\begin{align}
\Phi^{\rm TOA}_{e^+} (E)
= \frac{E^2}{(E+ \phi_{e^+})^2} \left[
\Phi^{\rm sec,IS}_{e^+}(E+\phi_{e^+}) + \Phi^{\rm source,IS}_{e^+}(E + \phi_{e^+}) 
\right]\,,
\end{align}
where the typical value of the parameter $\phi_{e^+}$ varies between $0.5$~GV and $1.3$~GV. The electron flux, on the other hand, is parametrized as:
\begin{align}
\Phi^{\rm TOA}_{e^-}(E) & = \frac{E^2}{(E+\phi_{e^-})^2} \left[ C_1(E+\phi_{e^-})^{-\gamma_1} + C_{2} (E+\phi_{e^-})^{-\gamma_2}  \left( 1+ \frac{E}{E_s}\right)^{-\gamma_3}\right]\,. \label{eq:phie}
\end{align}
We note that the parametrization of electron flux in Ref.~\cite{Ibarra:2013zia} is only suitable for the range of $E< 500$~GeV and the spectrum is not suppressed when $E > 1$~TeV.  To solve the problem, we have slightly modified the parametrization so that the DAMPE data at $E>2$~TeV can be accommodated.  Values of the parameters used for estimating the background cosmic-ray electron and positron fluxes in this work are given in Table~\ref{tab:inputs}.

\begin{table}[htp]
\begin{tabular}{cccccc} \hline \hline 
 ~~~$C_{e^+}$~~~ &  ~~~$\gamma_{e^+}$~~~ & ~~~$C_{s}$~~~ & ~~~$\gamma_{s}$~~~ & ~~~$\phi_{e^+}$~~~ &~~~$E_s$~~~   \\ \hline 
  50 & 4.0 & 2.4 & 2.757 & 1.3 &1100   \\ \hline 
  ~~~$C_1$~~~ & $\gamma_1$ & $C_2$ & $\gamma_2$   & $\phi_{e^-}$ & $\gamma_3$ \\ \hline 
  3340 & 3.9 & 18 & 2.575 & 1.35  & 1.98 \\ 
\hline\hline
\end{tabular}
\caption{Values of parameters for the interstellar positron and electron fluxes, where the units of various parameters are: $[C_{e^+, s, 1,2}]={\rm s^{-1} sr^{-1} m^{-2} GeV^{-1}}$, $[\phi_{e^\pm}]={\rm GV}$, and $[E_{s}]={\rm GeV}$.  }
\label{tab:inputs}
\end{table}

We show in Fig.~\ref{fig:spectrum} our result of the $E^3$-scaled CRE spectrum in the energy range of 1~GeV to 10~TeV.  The background contribution is given by the black curve, which is seen to fit well the AMS-02 data (pink)~\cite{Aguilar:2014fea} for $E \alt 60$~GeV, the Fermi-LAT data (purple)~\cite{Abdollahi:2017nat} for $10~\mbox{GeV} \alt E \alt 500$~GeV, the CALET data (cyan)~\cite{Adriani:2017efm} for $10~\mbox{GeV} \alt E \alt 3$~TeV, and the DAMPE data (orange) for $E \agt 120$~GeV except around the $1.5$-TeV peak\,
\footnote{ It is noted that for the energy range of 70~GeV $\alt E \alt$ 500~GeV, the DAMPE data are consistently higher than the AMS-02 data.  Such a discrepancy may be partially attributed to the absolute energy scale calibration in both experiments~\cite{Ambrosi:2017wek}.}.
The red curve shows the result after including the DM contribution from a subhalo at a distance of $0.1$~kpc from the Earth.  The blue dashed (green dotted) curve shows how the $e^+ e^-$ energy spectrum gets smeared out if the subhalo is located $0.3$~kpc (1~kpc) away.

{\it Summary --}
In view of the peak structure in the cosmic-ray electron-and-positron spectrum around 1.5~TeV as reported recently by the DAMPE Collaboration, we revisit a dark matter model as an extension of the Higgs triplet model.  The dark matter candidate with a mass of about 3~TeV is the gauge boson associated with a hidden $SU(2)_X$ symmetry that is broken to its $Z_3$ subgroup by a quadruplet scalar field.  The stability of the dark matter is ensured by the discrete gauge symmetry.  The coupling between the $SU(2)_X$ quadruplet and the Higgs triplet facilitates the pair annihilation of dark matter particles into the charged Higgs bosons with mass slightly less than the dark matter candidate.  The annihilation cross section enjoys a Breit-Wigner enhancement when we take the mediator mass to be about 6~TeV.  We show the parameter space that can explain the observed dark matter relic abundance at $2\sigma$ level.

For a sufficiently small triplet vacuum expectation value induced by that of the Standard Model Higgs doublet, the charged Higgs bosons preferentially decay into lepton pairs.  Using 100\% branching ratios to the $e^\pm e^\pm$ and $e^\pm \nu$ modes respectively for the doubly-charged and singly-charged Higgs bosons, we show that it is possible to explain the 1.5-TeV peak as a result of the charged Higgs boson decays.  Moreover, we consider that the putative signal comes from a nearby dark matter subhalo.  The $e^+ e^-$ flux spectrum is evaluated with both background and propagation effects taken into account.  Our result agrees well with the AMS-02 data in the lower energy regime and the DAMPE in the higher energy regime.  The subhalo is preferred to locate at a distance of $\sim 0.1$~kpc away from us.

Finally, we note that because of the assumed little mixing between the 125-GeV Higgs boson and 6-TeV dark matter annihilation mediator, the scattering cross section between dark matter and nucleons is negligibly small.  Hence, the model can readily evade the constraints from direct searches.

{\it Note Added:} After finishing this work, a similar analysis also appeared~\cite{Li:2017tmd}.  In that work, the authors also employ leptonic decays of the Higgs triplet fields to explain the DAMPE $e^+e^-$ excess but has a scalar dark matter candidate.

\section*{Acknowledgments}
This research was supported in part by the Ministry of Science and Technology of Taiwan under Grant Nos.\ MOST-106-2112-M-006-010-MY2 and MOST-104-2628-M-002-014-MY4.

\end{document}